\documentclass[journal]{IEEEtai}

\usepackage[colorlinks,urlcolor=blue,linkcolor=blue,citecolor=blue]{hyperref}

\usepackage{color,array}

\usepackage{graphicx}

\ifCLASSOPTIONcompsoc

\usepackage[caption=false,font=normalsize,labelfon
t=sf,textfont=sf]{subfig}
\else
\usepackage[caption=false,font=footnotesize]{subfig}

\fi

\begin{document}

\title{Objective Evaluation of Deep Uncertainty Predictions for COVID-19 Detection}

\author{Hamzeh Asgharnezhad, Afshar Shamsi, Roohallah Alizadehsani,\\ 
Abbas Khosravi, \IEEEmembership{Senior, IEEE}, Saeid Nahavandi, \IEEEmembership{Fellow, IEEE}, Zahra Alizadeh Sani, and Dipti Srinivasan, \IEEEmembership{Fellow, IEEE}

\thanks{This manuscript is created on December 22, 2020. This research was partially supported by the Australian Research Council's Discovery Projects funding scheme (project DP190102181).}

\thanks{H. Asgharnezhad and A. Shamsi are individual researchers, Tehran, Iran (e-mails: \{hamzeh.asgharnezhad@gmail.com, afshar.shamsi.j@gmail.com\}).}

\thanks{R. Alizadehsani, A. Khosravi, and S. Nahavandi are with the Institute for Intelligent Systems Research and Innovation (IISRI), Deakin University, Australia (e-mails: \{ralizadehsani, abbas.khosravi, saeid.nahavandi\}@deakin.edu.au).}

\thanks{Zahra A. Sani is with Omid hospital, Iran University of Medical Sciences, Tehran, Iran (e-mail: d.alizadeh.sani@gmail.com).}

\thanks{D. Srinivasan is with Department of Electrical and Computer Engineering, National University of Singapore (e-mail: dipti@nus.edu.sg).}}

\maketitle

\begin{abstract}
Deep neural networks (DNNs) have been widely applied for detecting COVID-19 in medical images. Existing studies mainly apply transfer learning and other data representation strategies to generate accurate point estimates. The generalization power of these networks is always questionable due to being developed using small datasets and failing to report their predictive confidence. Quantifying uncertainties associated with DNN predictions is a prerequisite for their trusted deployment in medical settings. Here we apply and evaluate three uncertainty quantification techniques for COVID-19 detection using chest X-Ray (CXR) images. The novel concept of uncertainty confusion matrix is proposed and new performance metrics for the objective evaluation of uncertainty estimates are introduced. Through comprehensive experiments, it is shown that networks pertained on CXR images outperform networks pretrained on natural image datasets such as ImageNet. Qualitatively and quantitatively evaluations also reveal that the predictive uncertainty estimates are statistically higher for erroneous predictions than correct predictions. Accordingly, uncertainty quantification methods are capable of flagging risky predictions with high uncertainty estimates. We also observe that ensemble methods more reliably capture uncertainties during the inference.

\end{abstract}

\begin{IEEEImpStatement}
DNN-based solutions for COVID-19 detection have been mainly proposed without any principled mechanism for risk mitigation. The focus of existing literature is mainly on generating single-valued predictions using pretrained DNNs. In this paper, we comprehensively apply and comparatively evaluate three uncertainty quantification techniques for COVID-19 detection using chest X-Ray images. The novel concept of uncertainty confusion matrix is proposed and new performance metrics for the objective evaluation of uncertainty estimates are introduced for the first time. Using these new uncertainty performance metrics, we quantitatively demonstrate where and when we could trust DNN predictions for COVID-19 detection from chest X-rays. It is important to note the proposed novel uncertainty evaluation metrics are generic and could be applied for evaluation of probabilistic forecasts in all classification problems.

\end{IEEEImpStatement}


\section{Introduction}

\IEEEPARstart{T}{he} COVID-19 pandemic has greatly increased the demand for fast and reliable screening of suspected cases. Real-time reverse transcription-polymerase chain reaction is the gold standard for the COVID019 detection. While this diagnostic test has a high accuracy and sensitivity, it is time consuming, resource intensive, and expensive. These shortcomings and the rising positivity rates has led to a real need for auxiliary diagnostic tools which are fast, affordable, and available at scale. Common radiology images such as CXR or computed tomography (CT) contain salient information and visual indexes correlated with the COVID-19 infections~\cite{zu2020coronavirus}. Accordingly, the detection can be inferred from these modalities of suspected individuals suffering from COVID-19 symptoms. 

Several studies have been conducted since the onset of COVID-19 pandemic to automate its detection from chest radiology images using artificial intelligence techniques \cite{esteva2017dermatologist, esteva2019guide}. Deep neural networks (DNNs) and transfer learning~\cite{minaee2020deep} have been widely applied for this purpose due to their promising human-level or super-human level performance in object recognition tasks \cite{shoeibi2020automated, lalmuanawma2020applications, shi2020review, rekha2020role}.

DNN-based solutions for COVID-19 detection have been mainly proposed without any principled mechanism for risk mitigation. The focus of existing literature is mainly on generating single-valued predictions using pretrained DNNs \cite{shoeibi2020automated}. Proposed solutions are evaluated by point prediction-based performance metrics such as accuracy, sensitivity, specificity, and area under receiver operating characteristic (AUC). It is important to note that the transition from normal to COVID-19 is not always clear-cut. Difficult-to-diagnosis cases from radiology images could even lead to disagreement between experienced medical doctors. In fact, studies have shown that radiologists disagree with their colleagues 25\% of the time and themselves 20\% of the time \cite{daniel2005toman}. As the model decision has a direct impact on patient's treatment, it is of critical importance to know how confident DNNs are about their predictions. This information could be used for identifying patients which may best benefit from a medical second opinion. DNNs flagging potentially erroneous predictions due to high uncertainty can be used to mimic the common practice of requesting a second opinion from another health practitioner in medial settings \cite{raghu2019direct}. Such an uncertainty-aware decision-making pipeline could greatly improve the overall diagnosis performance.

In this paper, we comprehensively and quantitatively investigate the competency of DNNs for generating reliable uncertainty estimates for COVID-19 diagnosis. We first check the impact of pretraining using ImageNet and CXR image datasets on the network performance. Then MC-dropout (MCD), ensemble, and ensemble MC-dropout (EMCD) are implanted for quantifying uncertainties associated with point predictions of DNNs. Motivated by \cite {mukhoti2018evaluating, subedar2019uncertainty}, we introduce novel performance metrics for the comprehensive and quantitative evaluation of uncertainty estimates. The uncertainty estimate evaluation is conducted in a similar manner to that of binary classification evaluation. Through experiments, we try to shed light on whether uncertainty quantification methods proposed in literature can provide high uncertainty for erroneous predictions. This investigation is done both qualitatively (visually) and quantitatively (using new uncertainty evaluation metrics). 
\begin{figure}[!t]
 \centerline{ \includegraphics[width = .46\textwidth]{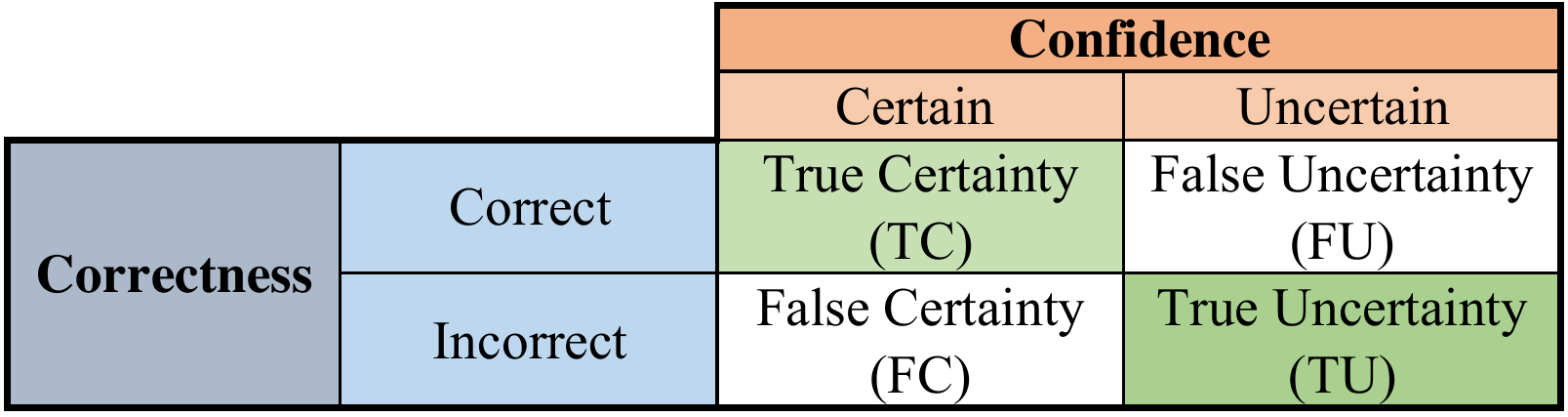}}
  \caption{The uncertainty confusion matrix}
  \label{Fig:UCM}
\end{figure}

\begin{figure*}
    \centering
    \begin{minipage}[b]{.44\textwidth}
    \subfloat[ImageNet]{\includegraphics[width=\textwidth]{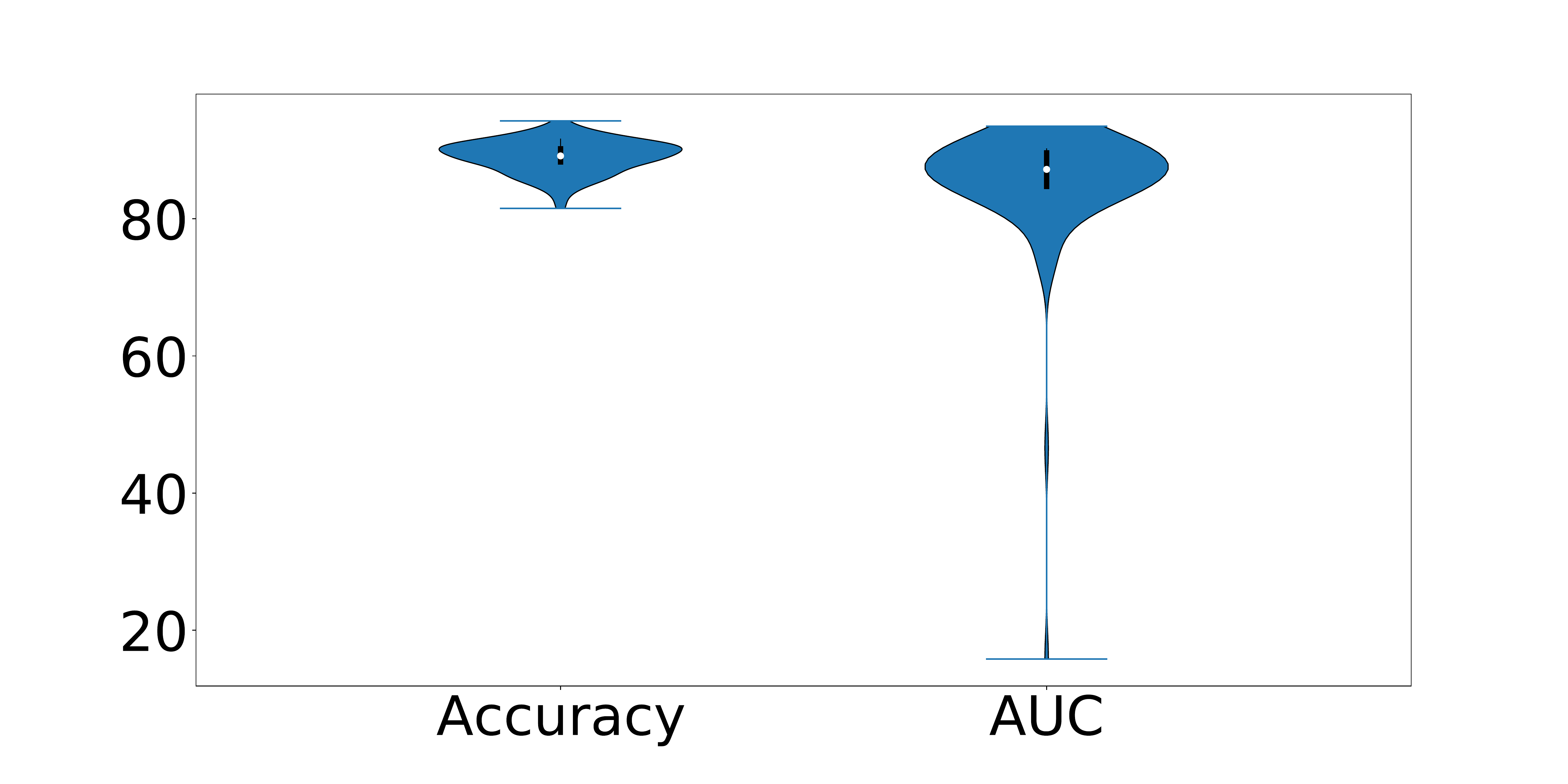}\label{Fig:TLcomp-ImageNet}}
    \end{minipage}\qquad
\begin{minipage}[b]{.44\textwidth}
    \subfloat[Pipeline]{\includegraphics[width=\textwidth]{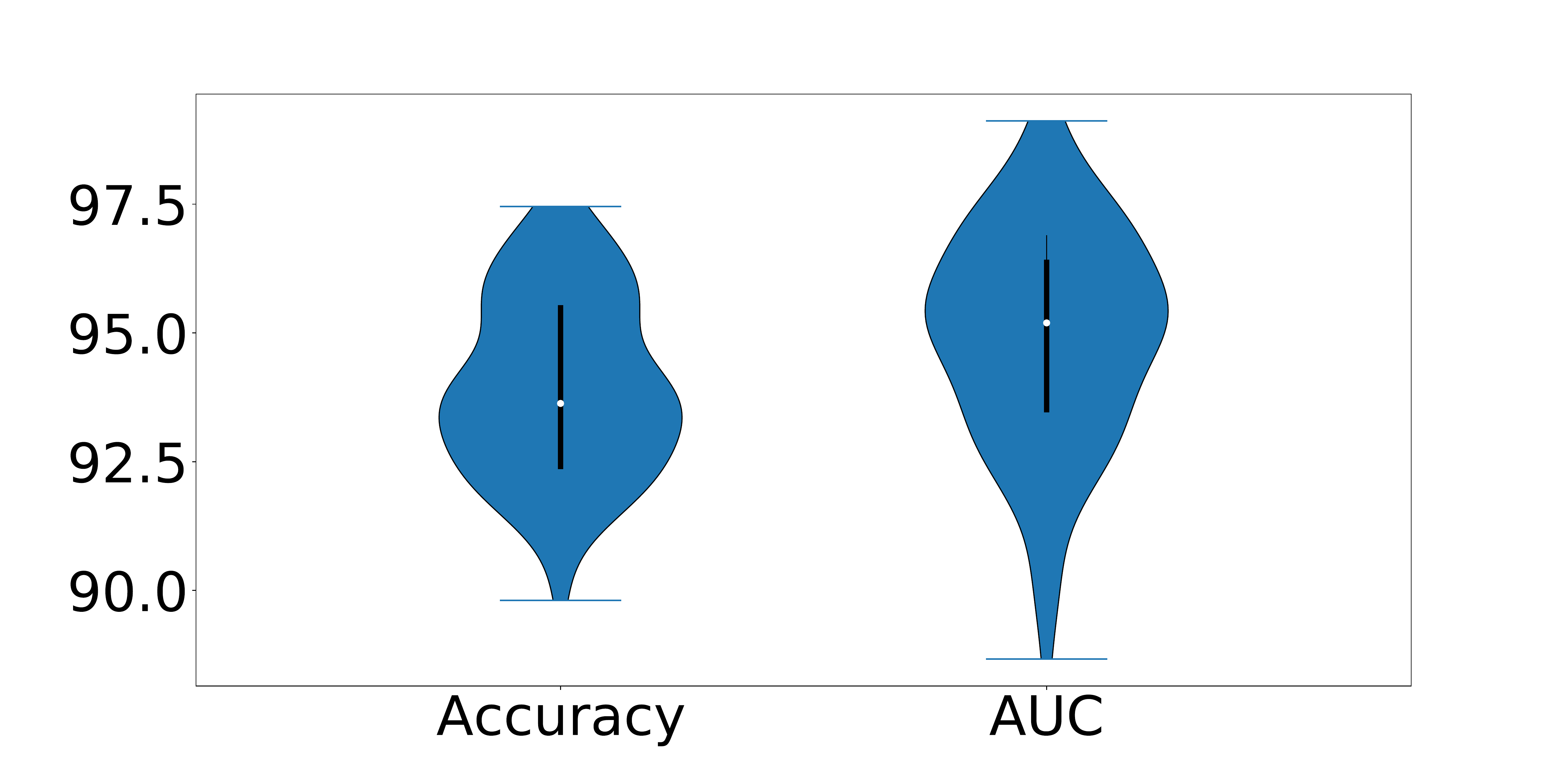}\label{Fig:TLcomp-ImageNet}}    
 \end{minipage}

    \caption{Accuracy and AUC (shown as a percentage) values for DenseNet121 pretrained using ImageNet and CXR datasets. The violin plot is obtained by training and measuring the network performance for 100 times.}
    \label{Fig:TLcomp-XRay}
\end{figure*}


The rest of this paper is organised as follows. Section \ref{Sec:RW} briefly reviews papers reporting applications of deep uncertainty quantification for COVID-19 diagnosis. Uncertainty quantification techniques are described in Section \ref{Sec:UQM}. Section \ref{Sec:PUE} introduces metrics for quantitative evaluation of predictive uncertainty estimates. The dataset and experiments are described in Section \ref{Sec:Dataset} and \ref{Sec:Exp} respectively. Section \ref{Sec:Res} reports conducted simulations and obtained results. Finally, section \ref{Sec:Conc} concludes the paper.

\section{Related Work}\label{Sec:RW}
Several methods have been proposed in recent years for enabling DNNs to encompass uncertainty and generate probabilistic predictions. Many of the proposed solutions are based on the Bayesian theory \cite{bernardo2009bayesian}. Several approximate methods have been proposed to address the intractability of the exact Bayesian inference due to its massive computational burden. This include but not limited to variational inference \cite{graves2011practical, blundell2015weight}, MCD \cite{gal2016dropout}, ensemble \cite{lakshminarayanan2017simple}. All these methods could be put in the two-step category of \textit{uncertainty via classification} as they first train the model (classifier here) and then postprocess predictions to generate an uncertainty score \cite{raghu2019direct}. There have been also attempts to generate uncertainty estimates without resorting to sampling methods. \textit{Direct uncertainty prediction} methods train DNNs to directly generate uncertainty scores or distribution parameters in one single round of scoring \cite{choi2018uncertainty, raghu2019direct, vanuncertainty}. Despite made progress in this field, reliable generation of uncertainty estimates is still an open question and subject to further investigation. 
\begin{figure}[t]
   \centering
   \begin{minipage}[b]{.44\textwidth}
      \subfloat[Certain CXR]{\includegraphics[width=0.98\columnwidth]{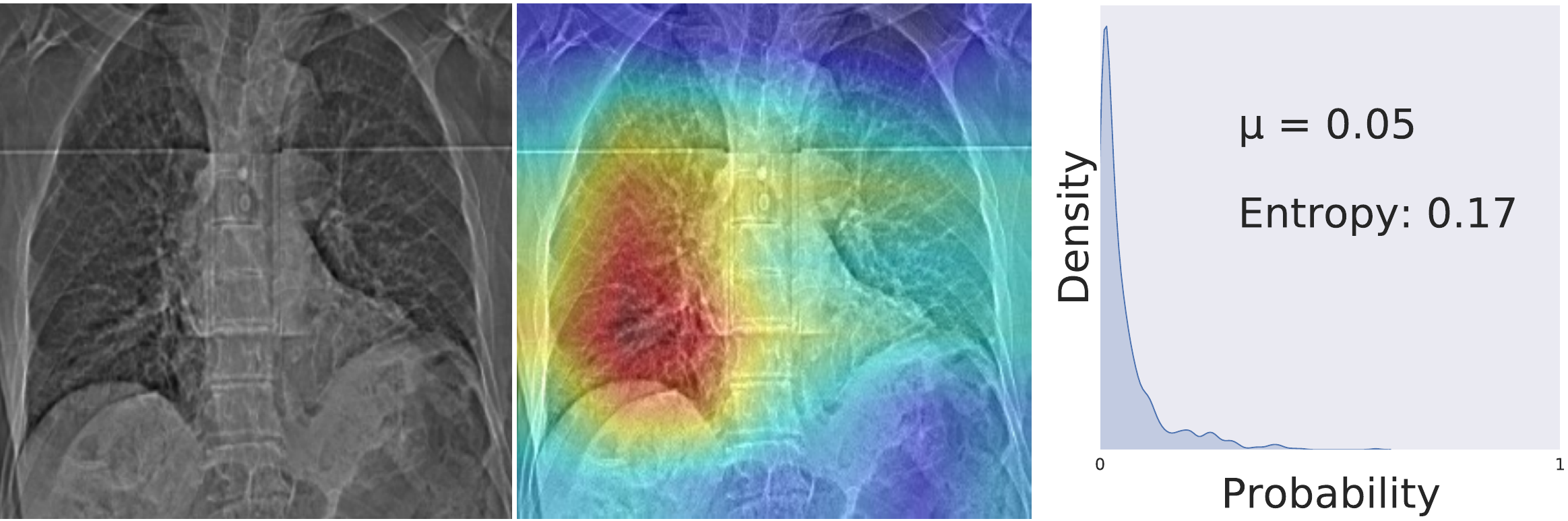}}
      \label{Fig:PostDist-Certain}
      \end{minipage}\qquad
    \begin{minipage}[b]{.44\textwidth}  
    \subfloat[Uncertain CXR]{\includegraphics[width=0.98\columnwidth]{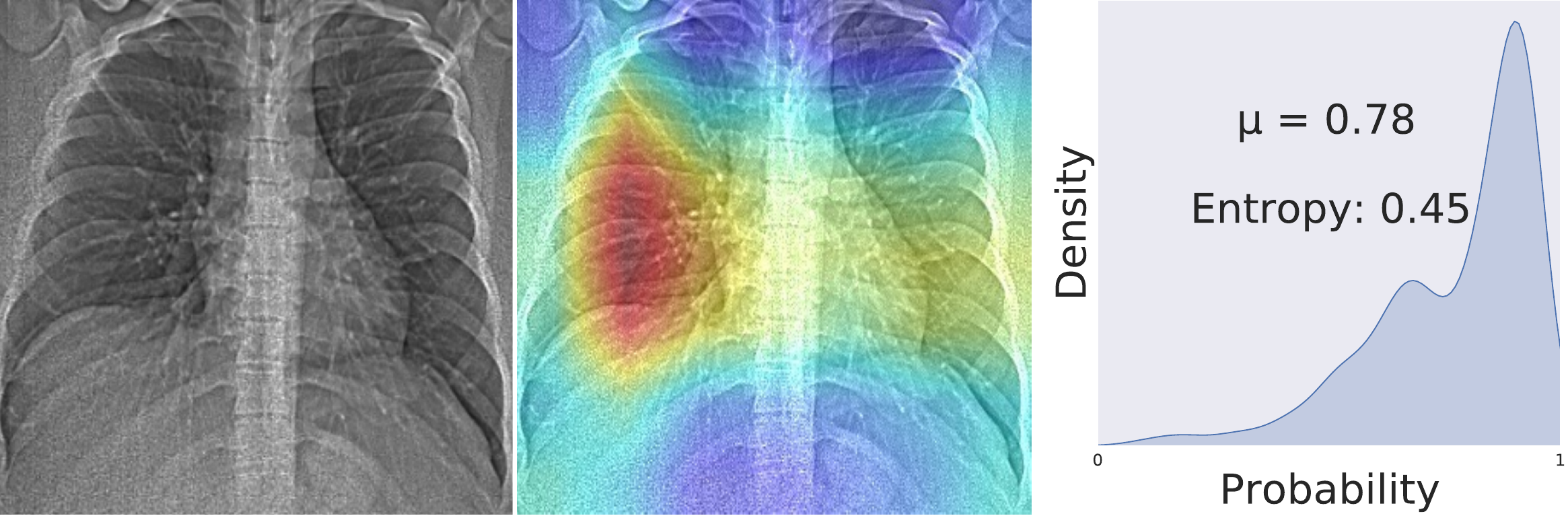}}
    \label{Fig:PostDist-Uncertain}
  \end{minipage}
  \caption{Two healthy (normal) images and their approximate predictive posterior distributions. These distributions, $p(covid | image)$, are estimated by the MCD algorithm. (a) correct and certain prediction, (b) incorrect and uncertain prediction. The middle plot in each row shows where DNNs look for making the decision.}
  \label{Fig:PostDist}
\end{figure}
\begin{figure*}[!h]
\centering
\begin{minipage}[b]{.3\textwidth}
  \subfloat[MCD]{\includegraphics[width=\textwidth]{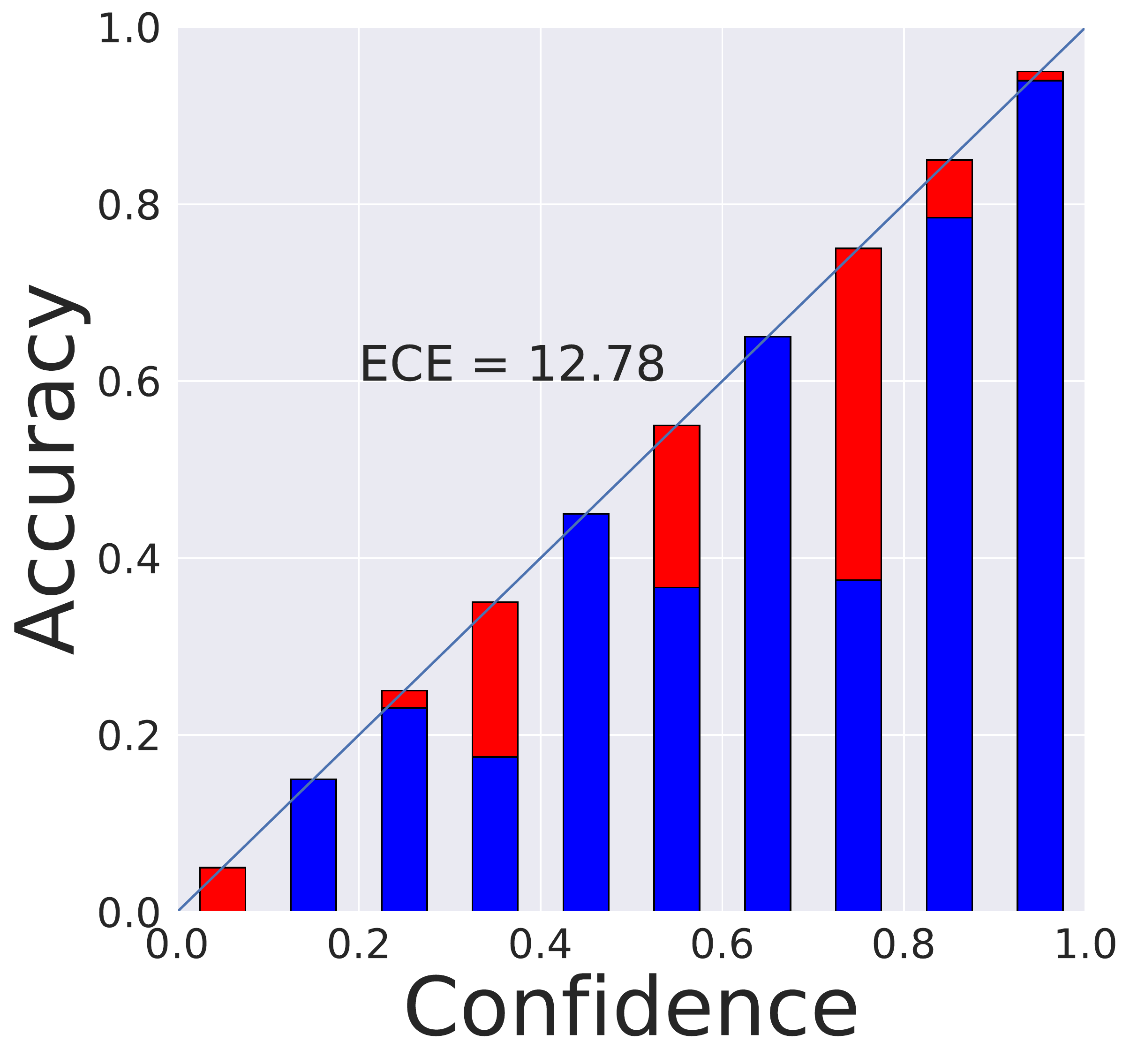}}
  \label{Fig:ECE of MCD for CT}
\end{minipage}\qquad
\begin{minipage}[b]{.3\textwidth}
  \subfloat[EMCD]{\includegraphics[width=\textwidth]{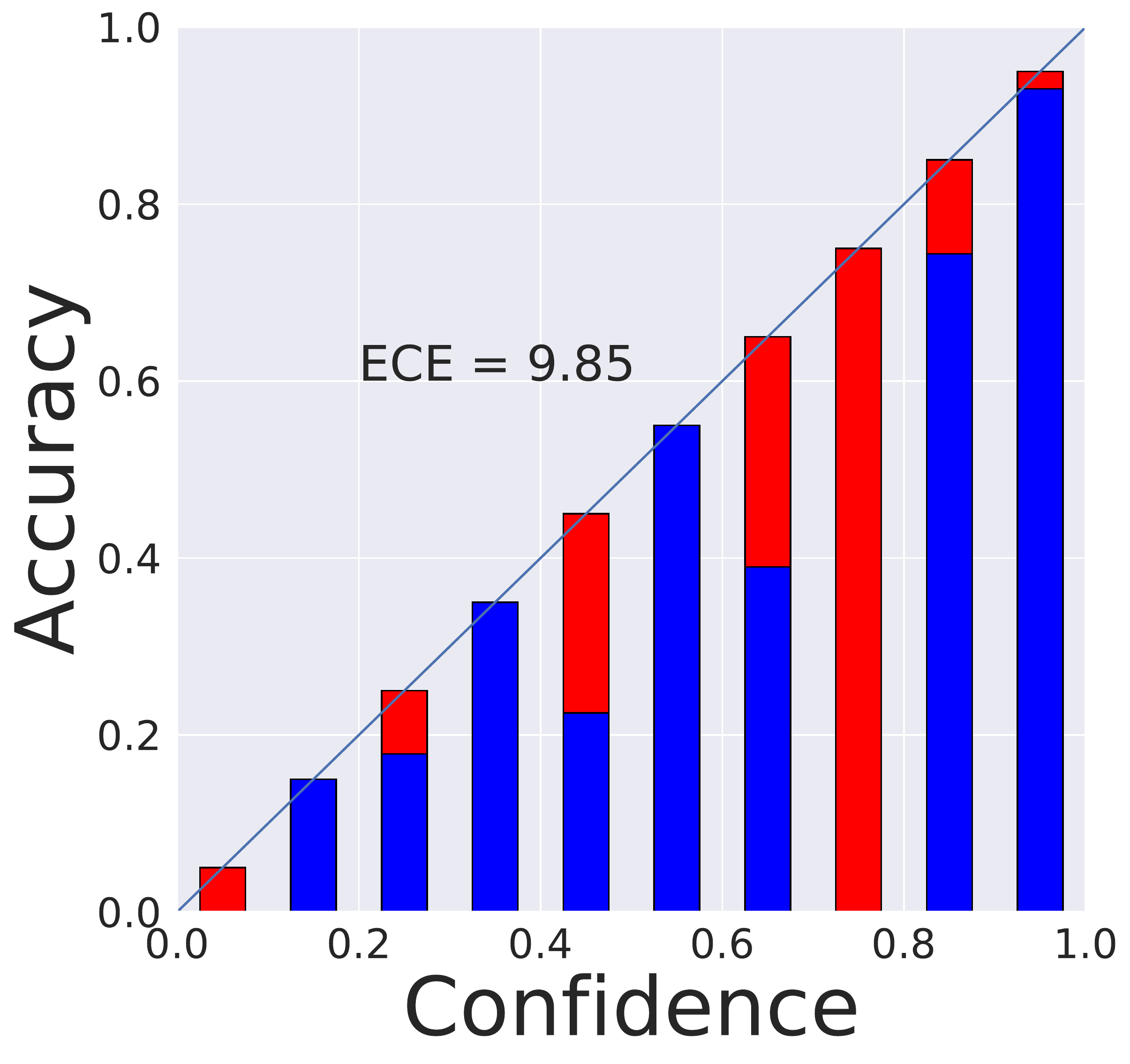}}
  \label{Fig:ECE of EMCD for CT}
\end{minipage}\qquad
\begin{minipage}[b]{.3\textwidth}
  \subfloat[Ensemble]{\includegraphics[width=\textwidth]{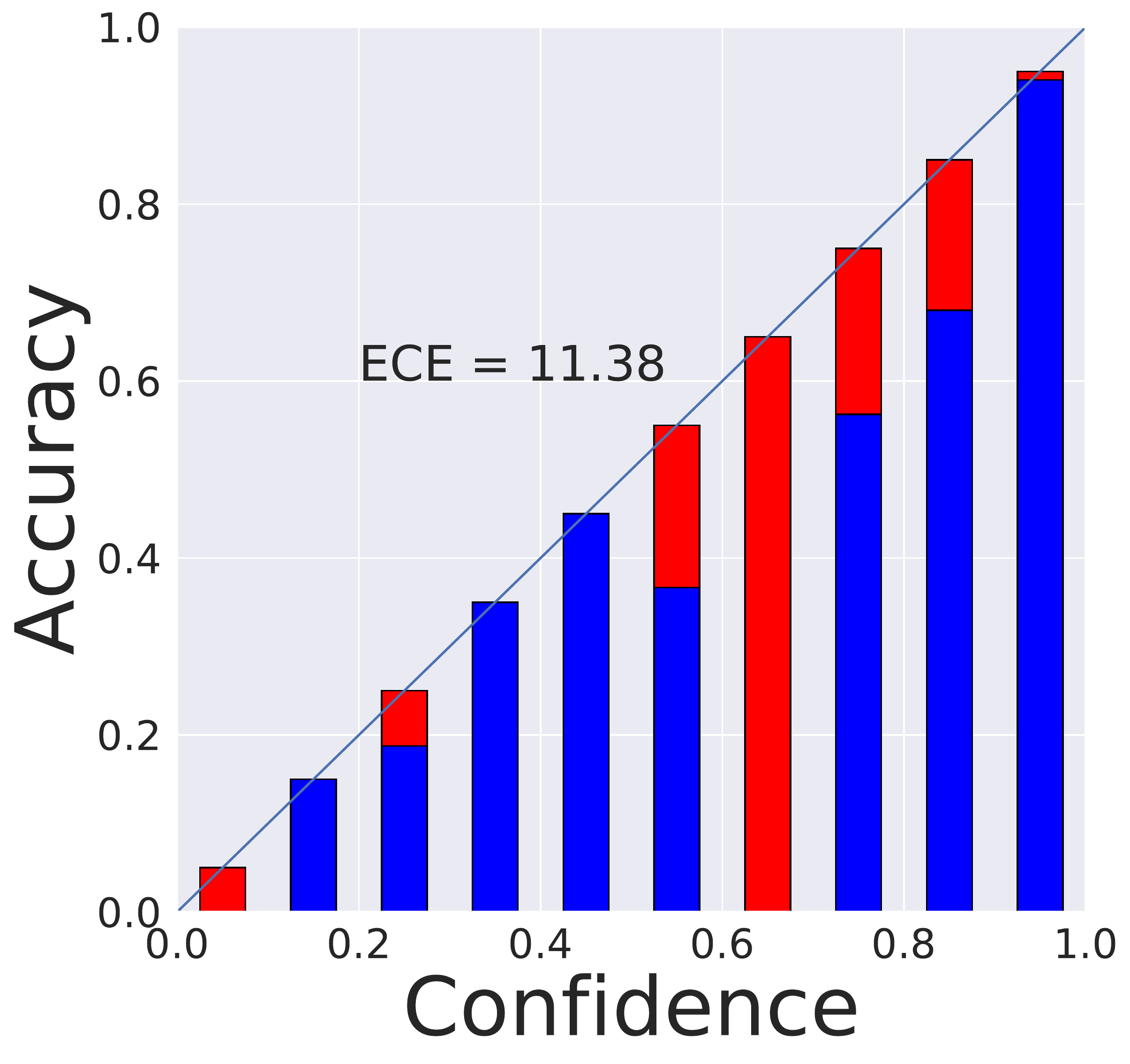}}
  \label{Fig:ECE of Ensemble for CT}
\end{minipage}
    \caption{The reliability diagrams (ECE plots) of the trained DNNs. MCD, EMCD, and Ensemble models all have a high ECE indicating miscalibration of generated probabilities.}
    \label{Fig:ECE}
\end{figure*}
There is an abundance of papers reporting applications of DNNs generating single-valued predictions for COVID-19 diagnosis from medical images \cite{shoeibi2020automated, lalmuanawma2020applications, shi2020review, rekha2020role}. \cite{shoeibi2020automated} provides a comprehensive review of medical imaging applications of DNNs for COVID-19. The input modalities are often chest X-ray and CT scan images which are processed using a wide variety of DNNs including pretrained networks for feature extraction, segmentation, and generative adversarial networks. While proposed solutions differ in terms of utilized networks and the task nature, they all focus on deterministic decisions generated by DNNs. This comprehensive review clearly shows that the literature is quite naive on applying deep uncertainty quantification techniques for processing COVID-19 datasets. Discussion about the reliability and confidence of proposed models has been often overlooked in these studies. 

There are a few rare studies reporting the importance of predictive uncertainty estimates for reliable COVID-19 detection from radiology images. The MC-Dropweights method \cite{ghoshal2019estimating} was used in ~\cite{ghoshal2020estimating} to estimate uncertainties associated with COVID-19 predictions. It was shown that there is a strong correlation between model uncertainty and prediction correctness. The paper clearly highlights that availability of estimated uncertainties could potentially alert radiologists on false predictions and will accelerate the acceptance of deep learning-based solution in clinical practice. Authors in ~\cite{jokandan2020uncertainty} apply four DNNs pretrained on ImageNet dataset to process CXR and CT images. Extracted features are used to develop an ensemble of neural networks for epistemic uncertainty quantification. Obtained results clearly highlight the need for uncertainty estimate to build trust in DNNS for COVID-19 detection. It is important to highlight that none of these studies provide a solid quantitative evaluation of uncertainty estimates generated by DNNs.

Authors in \cite{mallick50can} also propose a probabilistic generalization of the non-parametric KNN approach for developing a deep uncertainty-aware classifier. The proposed probabilistic neighbourhood component analysis method maps samples to probability distributions in a latent space and then minimizes a form of nearest-neighbour loss for developing classifiers. It is shown that the proposed method generates less overconfident predictions for out of distribution samples compared to common DNNs and Bayesian neural networks. Despite that, the paper does not provide any quantitative and qualitative evaluation about predictive uncertainty estimates for correctly classified and misclassified samples.


\section{Uncertainty Quantification Techniques}\label{Sec:UQM}

\subsection{MCD}
The most difficult part of the Bayesian network is finding the posterior distribution. This is often computationally intractable. One way to overcome this drawback is to use sampling methods. Gal~\cite{gal2016dropout} showed that MC samples of the posterior can be obtained by performing several stochastic forward passes at test time (keeping dropout on). The output posterior distribution could be approximated this way with minimum computational burden. The predictive mean ($\mu_{pred}$) of the model for a typical test input over MC iterations is estimated as below:

\begin{equation}
    \mu_{pred} \approx \frac{1}{T} \sum_t p(y = c | x, \hat\omega_t)
\end{equation}

\noindent where $x$ is the test input. $p(y = c | x, \hat\omega_t)$ is the probability that $y$ belongs to $c$ (the output of softmax), and $\hat\omega_t$ is the set of parameters of the model on the $t^{th}$ forward pass. $T$ is the number of MC iterations (forward passes). The variance of the final distribution is also called predictive uncertainty. As per \cite{gal2016dropout}, the predictive entropy (PE) can be treated as the uncertainty estimate generated by the trained model:

\begin{equation}\label{Eq:MC-Dropout-PE}
    PE = - \sum_c \mu_{pred} \log \mu_{pred}
\end{equation}

\noindent where $c$ ranges over both classes. The smaller the PE, the more confident the model about its predictions.


\subsection{Ensemble Bayesian Networks}
Ensemble networks are a group of networks working together for a specific task. Each network predicts a probability and the mean of probabilities will resemble the final predictive probability (posterior). The PE measure is also defined as~\cite{van2020simple}:

\begin{equation}
    \hat{p} (y|x)\ = \ \frac{1}{N}\ \sum\limits_{i=1}^N p_{\theta_{i}} (y|x) \label{eq:1}
\end{equation}

\begin{equation}
   PE \ = \ \sum\limits_{i=0}^C \hat{p}(y_i | x)\ log\ \hat{p}(y_i | x) \label{eq:2}
\end{equation}

\noindent where $\theta_i$ represents the set of parameters of $i_{th}$ network element, and $C$ ranges over two classes. The PE value is small when predictions from all individual networks are similar.

\subsection{EMCD}
A combination of Ensemble networks and MCD algorithms produces EMCD. Here the ensemble is consist of DNNs with different architectures. The evaluation of each network is done using the MCD algorithm by performing several stochastic forward passes. A single Gaussian distribution will be estimated by averaging all posterior probabilities. For PE metric, the algorithm is similar to the ensemble and just differs in the way of finding the posterior:

\begin{equation}
    \hat{p} (y|x) \approx \frac{1}{T} \sum_{t=1}^{T} p(\hat{y} | \hat{x},\hat\omega_t )
\end{equation}

\begin{equation}
   PE \ = \ \sum\limits_{i=0}^C \hat{p}(y_i | x)\ log\ \hat{p}(y_i | x) \label{eq:2}
\end{equation}

\noindent where $\hat\omega_t$ are the parameters of the model and $C$ ranges over both classes.

\section{Predictive Uncertainty Evaluation}\label{Sec:PUE}
Similar to the idea of confusion matrix, here we define quantitative performance metrics for predictive uncertainty estimates. In contrast to \cite{ghoshal2020estimating}, the purpose is to do an objective and quantitative evaluation of the predictive uncertainty estimates. Predictions are first compared with ground truth labels and put into two groups: correct and incorrect. Predictive uncertainty estimates are also compared with a threshold and cast into two groups: certain and uncertain. The combination of correctness and confidence groups results in four possible outcomes as shown in Fig. \ref{Fig:UCM}: (i) correct and certain indicated true certainty (TC), (ii) incorrect and uncertain indicated by true uncertainty (TU), correct and uncertain indicated by false uncertainty (FU), and (iv) incorrect and certain indicated by false certainty (FC). TC and TU are the diagonal and favourite outcomes. These correspond to TN and TP outcomes in the traditional confusion matrix respectively. FU is a fortunate outcome as an uncertain prediction is correct. FC is the worst outcome as the network has confidently made an incorrect prediction. 

According to these, we define multiple quantitative performance metrics to objectively quantify predictive uncertainty estimates:

\begin{itemize}
    \item Uncertainty sensitivity (USen): USen is calculated as the number of incorrect and uncertain predictions divided by the total number of incorrect predictions:
    
    \begin{equation}\label{Eq:USen}
        USen \ = \ \frac{TU}{TU + FC}
    \end{equation}
    
    \noindent USen or uncertainty recall (URec) corresponds to sensitivity (recall) or true positive rate of the conventional confusion matrix. USen is of paramount importance as it quantifies the power of the model to communicate its confidence in misclassified samples.
    
    \item Uncertainty Specificity (USpe): USpe is calculated as the number of correct and certain predictions (TC) divided by the total number of correct predictions: 
    
    \begin{equation}\label{Eq:USpe}
        USpe \ = \ \frac{TC}{TC + FU}
    \end{equation}
    
    USpe or correct certain ratio is similar to the specificity performance metric. 
    
    \item Uncertainty precision (UPre): UPrec is calculated as the number of incorrect and uncertain predictions divided by the total number of uncertain predictions:
    
    \begin{equation}\label{Eq:UPre}
        UPre \ = \ \frac{TU}{TU + FU}
    \end{equation}
    
    UPre has the same concept of precision in traditional binary classification.
    
    \item Uncertainty accuracy (UAcc): Similar to the accuracy of classifiers, the UAcc is calculated as the number of all diagonal outcomes divided by the total number of outcomes:
    
    \begin{equation}\label{Eq:UAcc}
        UAcc \ = \ \frac{TU + TC}{TU + TC + FU + FC}
    \end{equation}
    
    A reliable model will generate a high UAcc. 
    
\end{itemize}

The best USen, USpe, UPre, and UAcc values are one, whereas the worst are zero. It is always desirable to have these metrics as close as possible to one. Having USen and USpe, and UPre close to one means that the network is self-aware of what it knows and what it does not know. Such a network can tell us when the user can trust its predictions as it reliably gauges and communicates its lack of confidence (as captured in predictive uncertainty estimates).

\begin{figure*}[!t]
    \centering
 \begin{minipage}[b]{.3\textwidth}
    \subfloat[MCD]{\includegraphics[width=\textwidth]{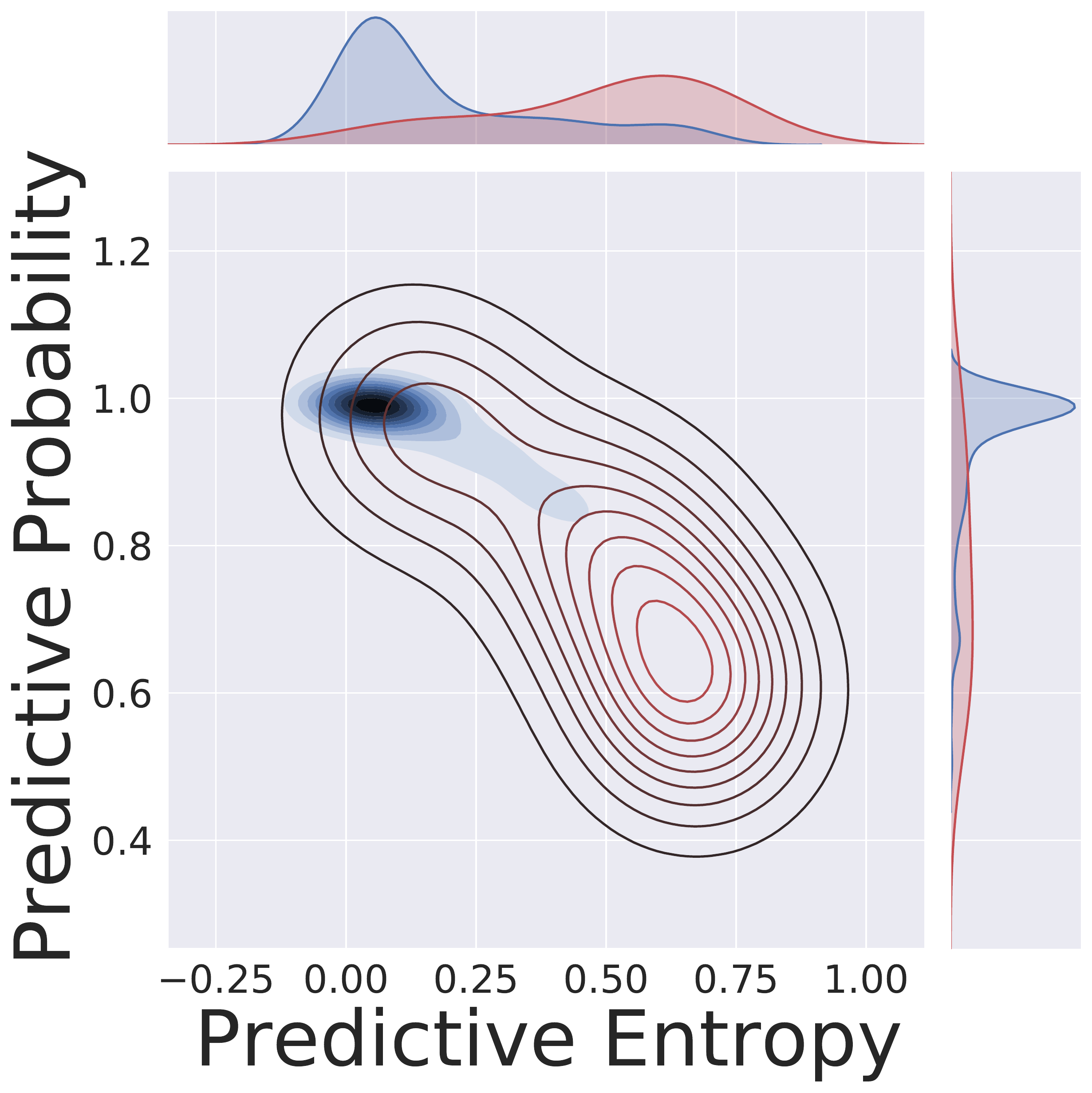}}
    \label{Fig:CT-MC-Dropout}
 \end{minipage}\qquad
 \begin{minipage}[b]{.3\textwidth} 
     \subfloat[EMCD]{\includegraphics[width=\textwidth]{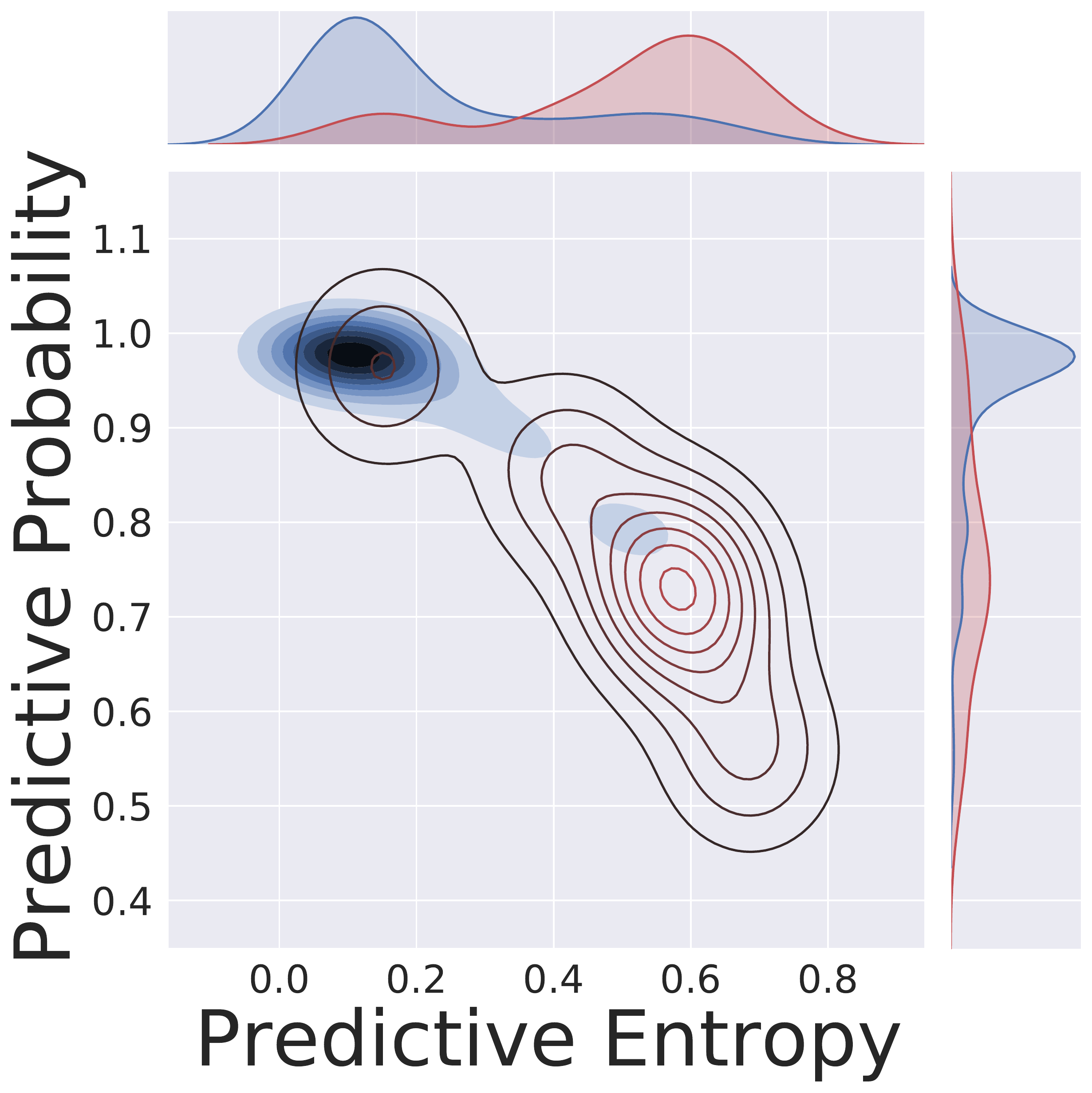}}
    \label{Fig:CT-Ensemble-MC-Dropout}
 \end{minipage}\qquad
\begin{minipage}[b]{.3\textwidth} 
    \subfloat[Ensemble]{\includegraphics[width=\textwidth]{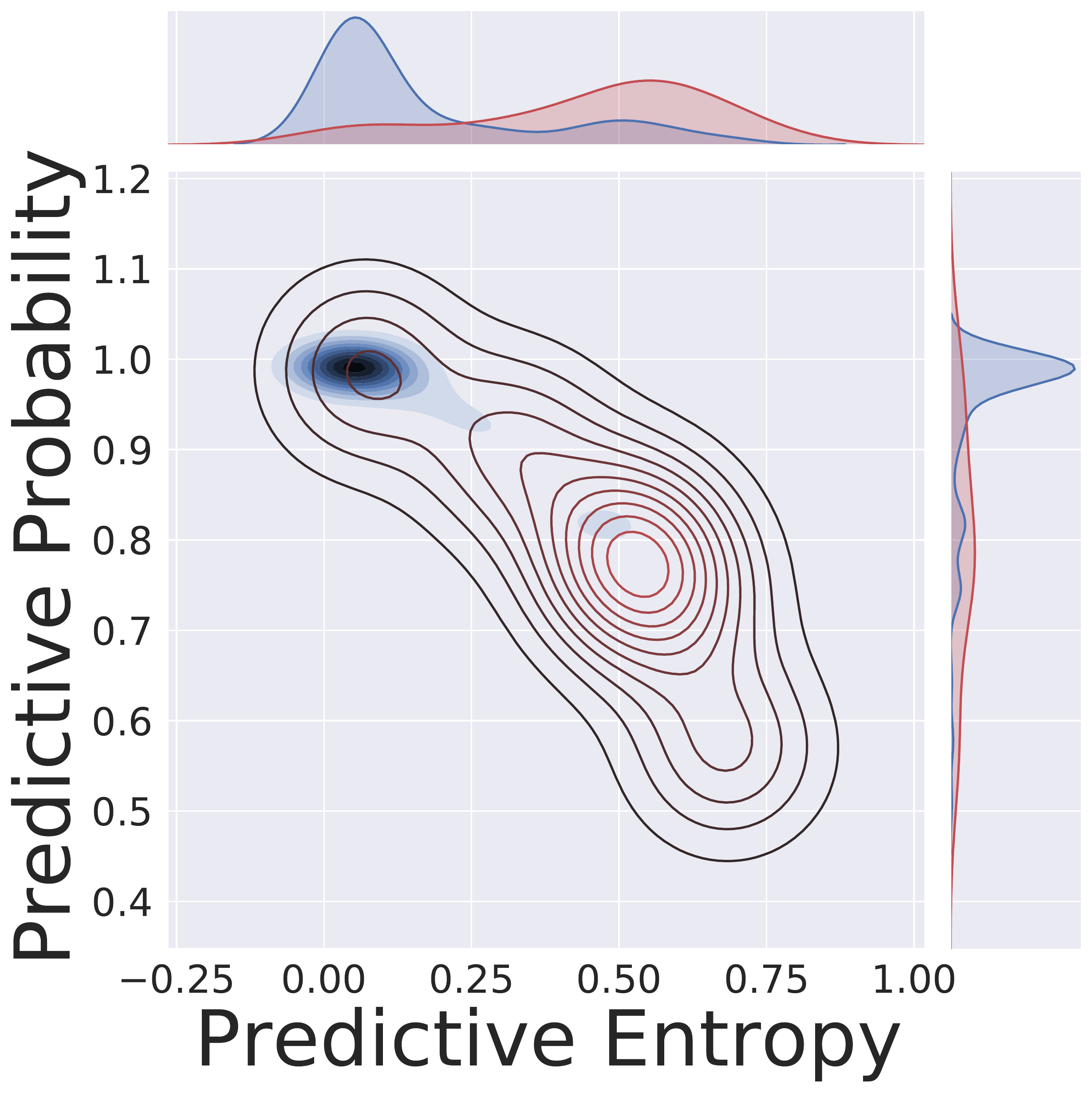}}
    \label{Fig:CT-Ensemble}
\end{minipage}
    
    \caption{Contour plots of predicted probabilities and uncertainty estimates (entropy) for three uncertainty quantification methods. Distribution of estimated predictive uncertainty estimates grouped by correctly classified and misclassified are shown on the top of plots. The best separation between two groups is obtained by ensemble methods.}
    \label{Fig:Contours}
\end{figure*}

\section{Dataset}\label{Sec:Dataset}
In this study, CXR images sourced from two databases are used for model training and testing. 
\subsection{Non-COVID Dataset}\label{SubSec:NCD}
Cohen et al~\cite{cohen2020predicting} developed a CXR image database by combining seven existing non-covid datasets: RSNA Pneumonia Challenge \cite{shih2019augmenting}, CheXpert - Stanford University \cite{irvin2019chexpert}, ChestX-ray8 - National Institutes of Health (NIH) \cite{wang2017chestx}, ChestX-ray8 - NIH with labels from Google \cite{majkowska2020chest}, MIMIC-CXR - MIT \cite{johnson2019mimic}, PadChest - University of Alicante \cite{bustos2020padchest}, and OpenI \cite{demner2016preparing}. This dataset is mainly used for training a DenseNet121 network which will be later on used as a pretrained network for uncertainty-aware COVID-19 detection.
\subsection{COVID-19 Dataset}\label{SubSec:CD}
The main COVID-19 dataset used in this study for model development and evaluation contains $522$ CXR images from 391 COVID-19 patients and 131 normal subjects. It is important to note that the normal class represents patients that did not have COVID-19. The term normal here does not imply that these patients do not have any emerging disease.

\section{Experiments}\label{Sec:Exp}
The main COVID-19 dataset has a limited number of images. This makes developing reliable DNNs from scratch impractical. To address this issue, we develop the deep model in a transfer learning setting \cite{minaee2020deep}. The common research practice is to pick an existing deep network pretrained on natural image datasets such as ImageNet and then finetune its weights on the medical images. It has been recently shown that this approach is not optimal for medical imaging \cite{raghu2019transfusion}. Motivated by these findings, we pretrain a DenseNet121 \cite{huang2017densely} using thousands of CXR images of the non-COVID datasets described in section \ref{SubSec:NCD}. 

The whole dataset is split to $75\% - 25\%$ between training and testing subsets. All images are resized to $224 \times 224$ and standardised before being fed to convolutional layers of DenseNet121. This results in $50,176$ convolutional features which are then processed by fully connected layers with a softmax on top of them. Relu activation function, $300$ epochs, and dropout rate of $0.25$ are used for model development using three uncertainty quantification techniques. The Adam algorithm with a learning rate of $0.001$ is applied to optimize the cross entropy loss function. For the MCD model, the number of neurons in three fully connected layers is set to $512$, $256$ and $64$ respectively. The ensemble model consists of 30 individual networks in which hidden layers are randomly chosen between two and three. Also the number of neurons in fully connected layers is randomly chosen between $(512, 1024)$, $(128, 512)$, $(8, 128)$ respectively. The ensemble MCD is designed similar to the ensemble. The only difference is that the evaluation of each network is done by the MCD algorithm.
\section{Simulations and Results}\label{Sec:Res}
We first compare the performance of DenseNet121 networks pretrained using ImageNet and CXR datasets. Fig. \ref{Fig:TLcomp-XRay} shows the violin plot of the accuracy and AUC performance metrics for these two networks trained and evaluated 100 times. Both performance metrics are greater and more consistent for networks pretrained using CXR datasets. A paired t-test is also run at 95\% confidence level to determine whether the mean of performance metrics are statistically different. The obtained p-values for accuracy and AUC are $10^{-39}$ and $10^{-19}$. As both values are much smaller than 0.05, it can be concluded that there are statistically significant differences in accuracy and AUC of these two models.

Results shown in Fig. \ref{Fig:TLcomp-XRay} indicate that the proposed pipeline does a promising job for COVID-19 detection from medical images. The model AUC and accuracy are $0.95 \pm 0.02$ and $93.94\% \pm 1.73\%$ which are in par or better than results reported in similar studies \cite{islam2020review}.
Before starting to analyze predictive uncertainty estimation results, we first check the calibration of predictions generated by DNNs. Fig. \ref{Fig:ECE} shows the expected calibration error (ECE) which is a plot of sample accuracy as a function of confidence \cite{guo2017calibration}. To calculate ECE, predictions are grouped in different bins (here $M$ bins) according to their confidence (the value of the max softmax output). The calibration error of each bin measures the difference between the fraction of correctly classified predictions (accuracy) and the mean of the probabilities (confidence). ECE is a weighted average of this error across all bins \cite{guo2017calibration}:

\begin{equation}\label{Eq:ECE}
    ECE = \sum_{m=1}^{M} \frac{|B_m|}{n} \left | acc(B_m) - conf(B_m) \right |
\end{equation}

\noindent where $ acc(B_m) $ and $ conf(B_m)$ are the accuracy and confidence in the m-th bin:
\begin{equation}\label{Eq:ECE-Acc}
    acc(B_m) = \sum \frac{1}{|B_m|} \textbf{1} \left (\hat{y_i} = y_i \right )
\end{equation}

\begin{equation}\label{Eq:ECE-Conf}
    conf(B_m) = \sum \frac{1}{|B_m|} p_i
\end{equation}

\noindent where $\textbf{1}(\cdot)$ is the indicator function.

In Fig. \ref{Fig:ECE}, the more the blue part deviation from the corresponding red parts, the less calibrated the neural network model. A perfectly calibrated model will generate the identity line in this chart. The current plot clearly shows that probabilities generated by three DNN families investigated in this paper are not calibrated. ECE values reported in Fig. \ref{Fig:ECE} indicate that all models overconfidently classify samples resulting in misleading outcomes. EMCD has the smallest ECE value of 9.85

\begin{figure*}
\centering
\begin{minipage}[b]{.22\textwidth}
    \subfloat[Unc. accuracy]{\includegraphics[width=\textwidth]{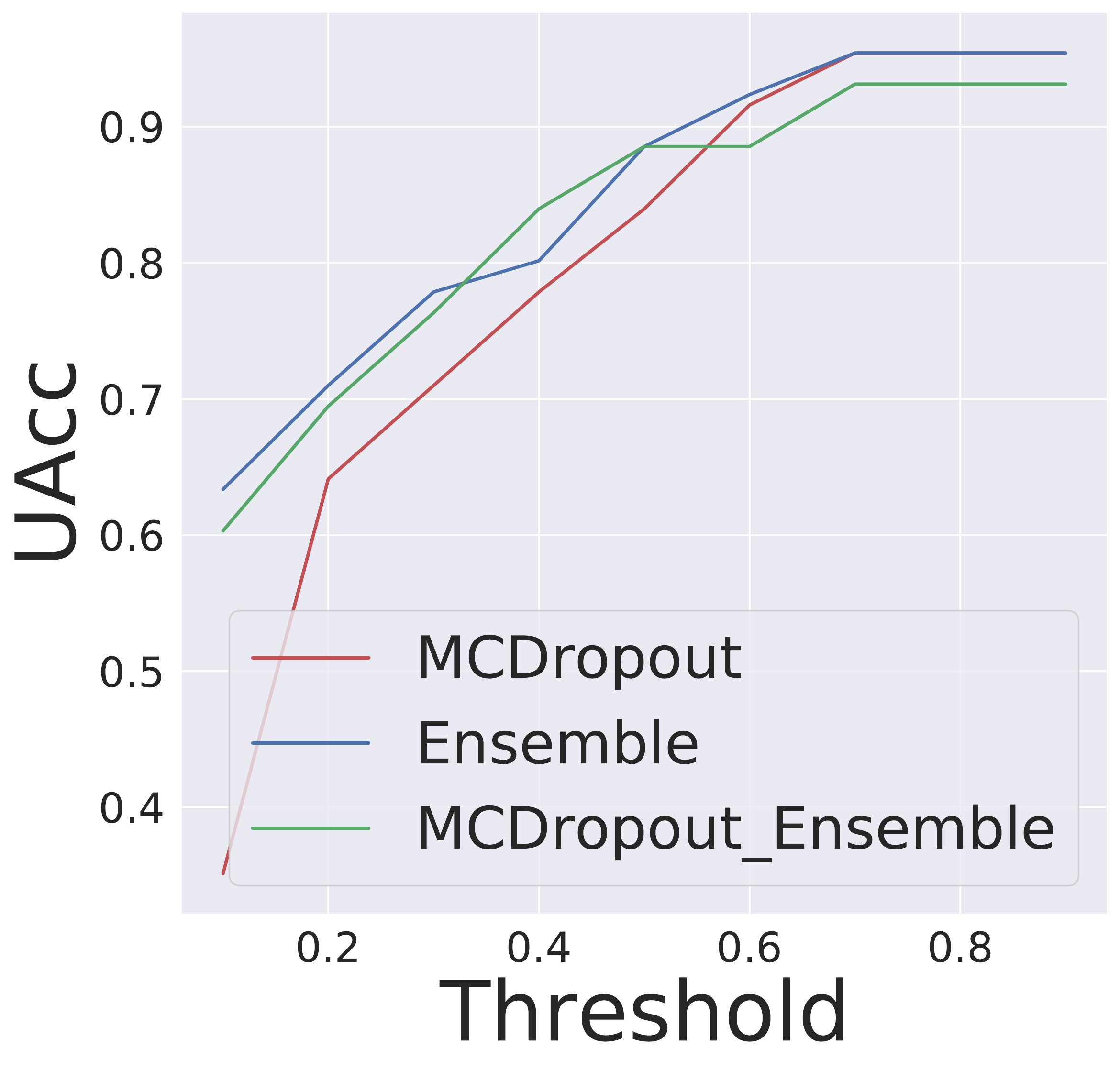}}
    \label{Fig:UAcc}
\end{minipage}\qquad
\begin{minipage}[b]{.22\textwidth}
    \subfloat[Unc. sensitivity]{\includegraphics[width=\textwidth]{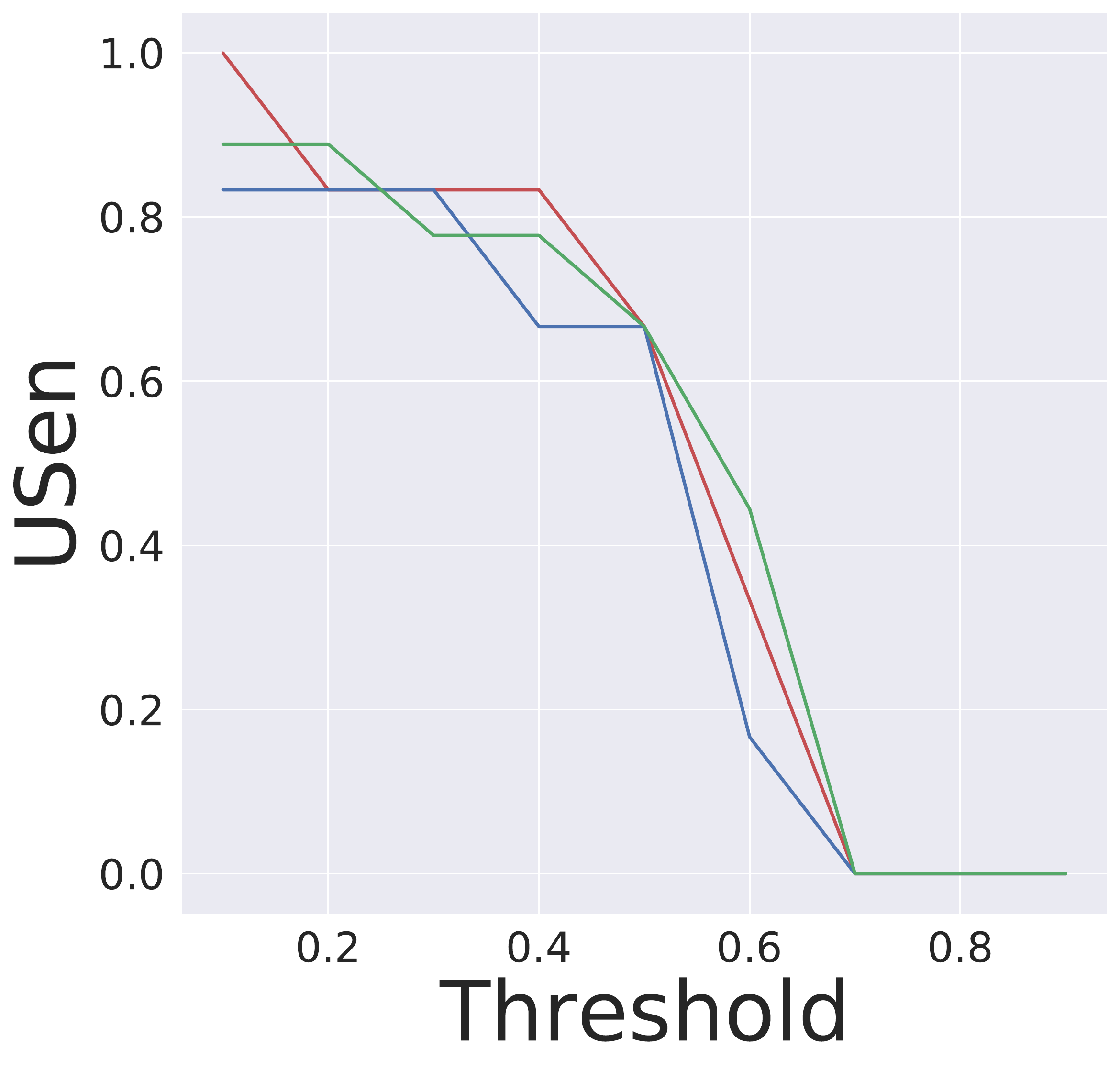}}
    \label{Fig:USen}
\end{minipage}\qquad
\begin{minipage}[b]{.22\textwidth}
    \subfloat[Unc. specificity]{\includegraphics[width=\textwidth]{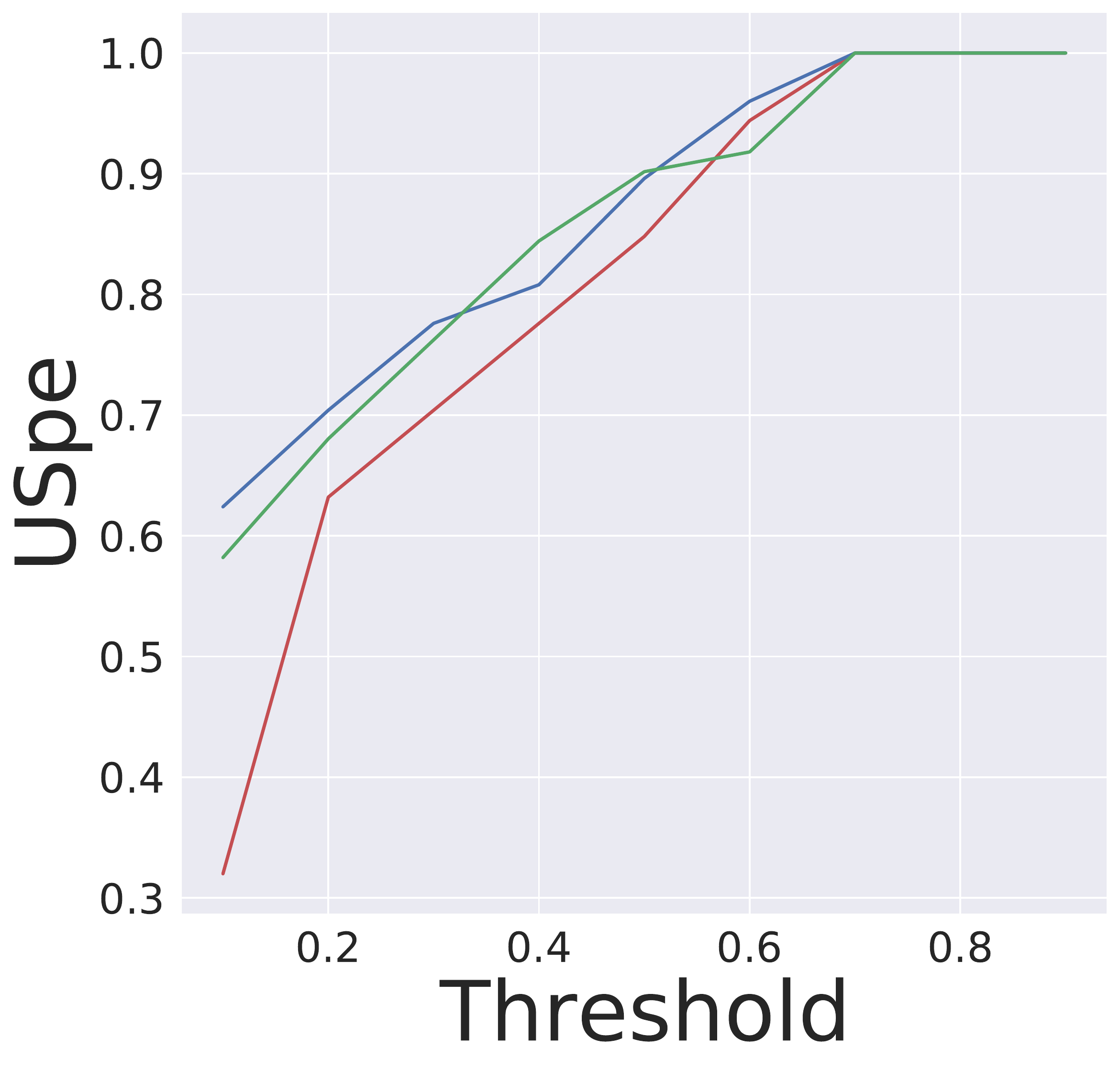}}
    \label{Fig:USpe}
\end{minipage}\qquad
\begin{minipage}[b]{.22\textwidth}
    \subfloat[Unc. precision]{\includegraphics[width=\textwidth]{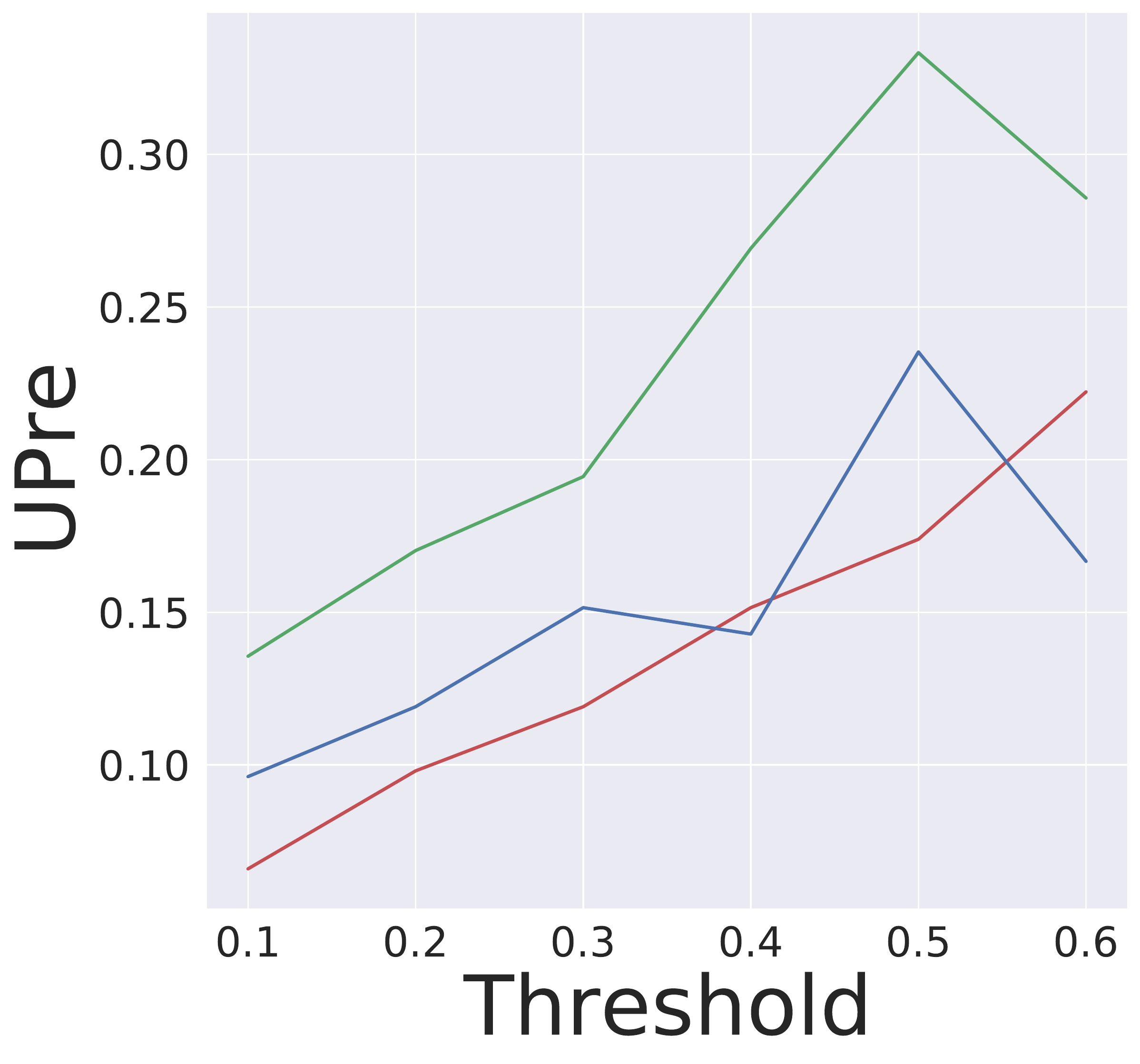}}
    \label{Fig:UPre}
\end{minipage}
    \caption{Quantitative evaluation of three uncertainty quantification techniques using performance metrics introduced in section \ref{Sec:PUE}. Uncertainty accuracy, sensitivity, specificity, and prediction are calulated for threshold values between 0.1 and 0.9.}
    \label{Fig:UncMetrics}
\end{figure*}

Fig. \ref{Fig:PostDist} displays the predictive posterior distribution for two non-COVID-19 (normal) CXR images obtained using MCD method (200 MC iterations). The class with the bigger softmax output for the distribution mean is reported as the predicted outcome. The uncertainty estimate associated with this outcome is also calculated using equation (\ref{Eq:MC-Dropout-PE}). The wider the predictive posterior distribution, the less confident the model. For the case shown in Fig. \ref{Fig:PostDist-Certain}, the predicted output is normal and the model is confident about its correct decision (a low predictive uncertainty estimate of 0.17). In contrast, the model prediction is wrong for the right CXR image in Fig. \ref{Fig:PostDist-Uncertain}. However, the model generates a wide posterior distribution resulting in a high predictive uncertainty estimate (0.45). This way, it communicates its lack of confidence in this specific prediction and says \textit{I do not know}. As the model is not confident about its prediction, the image could be sent to a medical expert for a \textit{second opinion} \cite{raghu2019direct}.

We then investigate the overall model ability in gauging and reporting its lack of confidence in its prediction. Fig. \ref{Fig:Contours} displays the contour plots of predicted probabilities and uncertainty estimates for three uncertainty quantification methods. The estimated marginal distributions of probabilities and uncertainty estimates are also shown on the sides of plots grouped by correctly classified and misclassified predictions. The plots clearly show that the centers of two groups are well apart from each other resulting in high accuracy. The estimated distribution for predicted probabilities of the correct group is much more compact compared to incorrect group. At the same time, the estimated uncertainties are higher for misclassified images. The visual inspection of three subplots indicates the uncertainty estimate mean for erroneous predictions is on the right of the uncertainty estimate mean for correct predictions. This qualitative investigation highlights the model capability in gauging and communicating its confidence (or lack of confidence) in generated predictions. This finding is of paramount practical importance as reliable uncertainty estimates provide additional valuable information to predicted probabilities. These could be used to flag uncertain predictions and request a second opinion by a medical expert. 

We also comprehensively evaluate predictive uncertainty estimates using performance metrics introduced in Section \ref{Sec:PUE}. Fig. \ref{Fig:UncMetrics} displays uncertainty accuracy (UAcc), uncertainty sensitivity (USen), uncertainty specificity (USpe), and uncertainty precision (UPre) calculated for uncertainty thresholds between 0.1 to 0.9. UAcc, USpe, and UPre are positively correlated with the uncertainty threshold. This correlation is negative for USen. UAcc, USen, and USpe all achieve values close to one for a wide range of thresholds. Achieving a high USen means that all three uncertainty quantification methods are able to flag incorrect predictions with high uncertainty. These methods are quite capable of flagging erroneous predictions for further investigation. None of the uncertainty quantification methods achieves a high UPre close to one. This is because there are many correct predictions that have a high uncertainty (FU). This can be observed in the long tail of the estimated distributions of predictive uncertainties in Fig. \ref{Fig:Contours} (top side plots). It is also important to note that the number of correctly classified images is much greater than the number of misclassified images. This makes the number of TU predictions always much smaller than FU resulting in a low Upre. This is an expected pattern for models with high accuracy.

UAcc, USpe, and Upre achieve their maximum values for thresholds close to one. Thresholds close to zero lead to the highest values for USpe. Selecting the best uncertainty threshold value depends on the users' preferences, e.g., sensitivity vs. specificity. Setting it to 0.3 results in a good trade off between four uncertainty performance metrics for three uncertainty quantification methods. Table. \ref{Tab:UncMetrics} reports uncertainty performance metrics for all methods. UAcc for the MCD method is 71.2\% which is much smaller than UAcc for both ensemble methods. The ensemble method achieves the highest UAcc amongst uncertainty quantification methods and its UAcc (77.8\%) is slightly superior to that of EMCD (76.3\%). The same pattern holds for other performance metrics of three methods.

\begin{table}[!t]
\centering
\caption{Uncertainty performance metrics for the specific threshold of 0.3 for three uncertainty quantification techniques.}\label{Tab:UncMetrics}
\resizebox{\columnwidth}{!}{%
\begin{tabular}{llllll}
        \hline
        UQ Method           & UAcc   & USen  & USpe   & UPre \\
        \hline
        MCD          & 71.2\% & 0.833  & 0.704   & 0.119    \\
        EMCD         & 76.3\% & 0.777  & 0.762   & 0.194    \\
        Ensemble     & 77.8\% & 0.833  & 0.776   & 0.151    \\
        \hline
    \end{tabular}
}
\end{table}

\section{Conclusion}\label{Sec:Conc}
In this paper, we investigate the competency of deep uncertainty quantification techniques for the task of COVID-19 detection from CXR images. A novel confusion matrix and multiple performance metrics for the evaluation of predictive uncertainty estimates are introduced. Our investigations reveal that deep learning models pretrained using medical imaging datasets outperform models pretrained using natural datasets such as ImageNet. Through comprehensive evaluation, we also find that ensemble methods better capture uncertainties associated with their predictions resulting in more trustworthy diagnosis solutions. The proposed uncertainty confusion matrix also shows that uncertainty quantification methods achieve high uncertainty sensitivity and specificity. However, they often fail at producing uncertainty estimates resulting in high precision.

There are many rooms for improving the proposed uncertainty evaluation metrics and its application for DNN development. For future work, we will include the proposed uncertainty evaluation metrics as the loss function in the process of training DNNs. This will lead to networks that are optimized based on performance metrics of both point predictions and uncertainty estimates.


\ifCLASSOPTIONcaptionsoff
  \newpage
\fi


\begin{thebibliography}{10}

\bibitem{zu2020coronavirus}
Z.~Y. Zu, M.~D. Jiang, P.~P. Xu, W.~Chen, Q.~Q. Ni, G.~M. Lu, and L.~J. Zhang,
  ``Coronavirus disease 2019 (covid-19): a perspective from china,'' {\em
  Radiology}, p.~200490, 2020.

\bibitem{esteva2017dermatologist}
A.~Esteva, B.~Kuprel, R.~A. Novoa, J.~Ko, S.~M. Swetter, H.~M. Blau, and
  S.~Thrun, ``Dermatologist-level classification of skin cancer with deep
  neural networks,'' {\em nature}, vol.~542, no.~7639, pp.~115--118, 2017.

\bibitem{esteva2019guide}
A.~Esteva, A.~Robicquet, B.~Ramsundar, V.~Kuleshov, M.~DePristo, K.~Chou,
  C.~Cui, G.~Corrado, S.~Thrun, and J.~Dean, ``A guide to deep learning in
  healthcare,'' {\em Nature medicine}, vol.~25, no.~1, pp.~24--29, 2019.

\bibitem{minaee2020deep}
S.~Minaee, R.~Kafieh, M.~Sonka, S.~Yazdani, and G.~J. Soufi, ``Deep-covid:
  Predicting covid-19 from chest x-ray images using deep transfer learning,''
  {\em arXiv preprint arXiv:2004.09363}, 2020.

\bibitem{shoeibi2020automated}
A.~Shoeibi, M.~Khodatars, R.~Alizadehsani, N.~Ghassemi, M.~Jafari, P.~Moridian,
  A.~Khadem, D.~Sadeghi, S.~Hussain, A.~Zare, {\em et~al.}, ``Automated
  detection and forecasting of covid-19 using deep learning techniques: A
  review,'' {\em arXiv preprint arXiv:2007.10785}, 2020.

\bibitem{lalmuanawma2020applications}
S.~Lalmuanawma, J.~Hussain, and L.~Chhakchhuak, ``Applications of machine
  learning and artificial intelligence for covid-19 (sars-cov-2) pandemic: A
  review,'' {\em Chaos, Solitons \& Fractals}, p.~110059, 2020.

\bibitem{shi2020review}
F.~Shi, J.~Wang, J.~Shi, Z.~Wu, Q.~Wang, Z.~Tang, K.~He, Y.~Shi, and D.~Shen,
  ``Review of artificial intelligence techniques in imaging data acquisition,
  segmentation and diagnosis for covid-19,'' {\em IEEE reviews in biomedical
  engineering}, 2020.

\bibitem{rekha2020role}
S.~rekha Hanumanthu, ``Role of intelligent computing in covid-19 prognosis: A
  state-of-the-art review,'' {\em Chaos, Solitons \& Fractals}, p.~109947,
  2020.

\bibitem{daniel2005toman}
T.~M. Daniel, ``Toman’s tuberculosis. case detection, treatment, and
  monitoring. questions and answers,'' {\em The American Journal of Tropical
  Medicine and Hygiene}, vol.~73, no.~1, pp.~229--229, 2005.

\bibitem{raghu2019direct}
M.~Raghu, K.~Blumer, R.~Sayres, Z.~Obermeyer, B.~Kleinberg, S.~Mullainathan,
  and J.~Kleinberg, ``Direct uncertainty prediction for medical second
  opinions,'' in {\em International Conference on Machine Learning},
  pp.~5281--5290, 2019.

\bibitem{mukhoti2018evaluating}
J.~Mukhoti and Y.~Gal, ``Evaluating bayesian deep learning methods for semantic
  segmentation,'' {\em arXiv preprint arXiv:1811.12709}, 2018.

\bibitem{subedar2019uncertainty}
M.~Subedar, R.~Krishnan, P.~L. Meyer, O.~Tickoo, and J.~Huang,
  ``Uncertainty-aware audiovisual activity recognition using deep bayesian
  variational inference,'' in {\em Proceedings of the IEEE International
  Conference on Computer Vision}, pp.~6301--6310, 2019.

\bibitem{bernardo2009bayesian}
J.~M. Bernardo and A.~F. Smith, {\em Bayesian theory}, vol.~405.
\newblock John Wiley \& Sons, 2009.

\bibitem{graves2011practical}
A.~Graves, ``Practical variational inference for neural networks,'' in {\em
  Advances in neural information processing systems}, pp.~2348--2356, 2011.

\bibitem{blundell2015weight}
C.~Blundell, J.~Cornebise, K.~Kavukcuoglu, and D.~Wierstra, ``Weight
  uncertainty in neural networks,'' {\em arXiv preprint arXiv:1505.05424},
  2015.

\bibitem{gal2016dropout}
Y.~Gal and Z.~Ghahramani, ``Dropout as a bayesian approximation: Representing
  model uncertainty in deep learning,'' in {\em international conference on
  machine learning}, pp.~1050--1059, 2016.

\bibitem{lakshminarayanan2017simple}
B.~Lakshminarayanan, A.~Pritzel, and C.~Blundell, ``Simple and scalable
  predictive uncertainty estimation using deep ensembles,'' in {\em Advances in
  neural information processing systems}, pp.~6402--6413, 2017.

\bibitem{choi2018uncertainty}
S.~Choi, K.~Lee, S.~Lim, and S.~Oh, ``Uncertainty-aware learning from
  demonstration using mixture density networks with sampling-free variance
  modeling,'' in {\em 2018 IEEE International Conference on Robotics and
  Automation (ICRA)}, pp.~6915--6922, IEEE, 2018.

\bibitem{vanuncertainty}
J.~van Amersfoort, L.~Smith, Y.~W. Teh, and Y.~Gal, ``Uncertainty estimation
  using a single deep deterministic neural network,''

\bibitem{ghoshal2019estimating}
B.~Ghoshal, A.~Tucker, B.~Sanghera, and W.~L. Wong, ``Estimating uncertainty in
  deep learning for reporting confidence to clinicians when segmenting nuclei
  image data,'' in {\em 2019 IEEE 32nd International Symposium on
  Computer-Based Medical Systems (CBMS)}, pp.~318--324, IEEE, 2019.

\bibitem{ghoshal2020estimating}
B.~Ghoshal and A.~Tucker, ``Estimating uncertainty and interpretability in deep
  learning for coronavirus (covid-19) detection,'' {\em arXiv preprint
  arXiv:2003.10769}, 2020.

\bibitem{jokandan2020uncertainty}
A.~S. Jokandan, H.~Asgharnezhad, S.~S. Jokandan, A.~Khosravi, P.~M. Kebria,
  D.~Nahavandi, S.~Nahavandi, and D.~Srinivasan, ``An uncertainty-aware
  transfer learning-based framework for covid-19 diagnosis,'' {\em arXiv
  preprint arXiv:2007.14846}, 2020.

\bibitem{mallick50can}
A.~Mallick, C.~Dwivedi, B.~Kailkhura, G.~Joshi, and T.~Y.-J. Han, ``Can your ai
  differentiate cats from covid-19? sample efficient uncertainty estimation for
  deep learning safety,'' {\em choice}, vol.~50, p.~6.

\bibitem{van2020simple}
J.~van Amersfoort, L.~Smith, Y.~W. Teh, and Y.~Gal, ``Simple and scalable
  epistemic uncertainty estimation using a single deep deterministic neural
  network,'' {\em arXiv preprint arXiv:2003.02037}, 2020.

\bibitem{cohen2020predicting}
J.~P. Cohen, L.~Dao, P.~Morrison, K.~Roth, Y.~Bengio, B.~Shen, A.~Abbasi,
  M.~Hoshmand-Kochi, M.~Ghassemi, H.~Li, {\em et~al.}, ``Predicting covid-19
  pneumonia severity on chest x-ray with deep learning,'' {\em arXiv preprint
  arXiv:2005.11856}, 2020.

\bibitem{shih2019augmenting}
G.~Shih, C.~C. Wu, S.~S. Halabi, M.~D. Kohli, L.~M. Prevedello, T.~S. Cook,
  A.~Sharma, J.~K. Amorosa, V.~Arteaga, M.~Galperin-Aizenberg, {\em et~al.},
  ``Augmenting the national institutes of health chest radiograph dataset with
  expert annotations of possible pneumonia,'' {\em Radiology: Artificial
  Intelligence}, vol.~1, no.~1, p.~e180041, 2019.

\bibitem{irvin2019chexpert}
J.~Irvin, P.~Rajpurkar, M.~Ko, Y.~Yu, S.~Ciurea-Ilcus, C.~Chute, H.~Marklund,
  B.~Haghgoo, R.~Ball, K.~Shpanskaya, {\em et~al.}, ``Chexpert: A large chest
  radiograph dataset with uncertainty labels and expert comparison,'' in {\em
  Proceedings of the AAAI Conference on Artificial Intelligence}, vol.~33,
  pp.~590--597, 2019.

\bibitem{wang2017chestx}
X.~Wang, Y.~Peng, L.~Lu, Z.~Lu, M.~Bagheri, and R.~M. Summers, ``Chestx-ray8:
  Hospital-scale chest x-ray database and benchmarks on weakly-supervised
  classification and localization of common thorax diseases,'' in {\em
  Proceedings of the IEEE conference on computer vision and pattern
  recognition}, pp.~2097--2106, 2017.

\bibitem{majkowska2020chest}
A.~Majkowska, S.~Mittal, D.~F. Steiner, J.~J. Reicher, S.~M. McKinney, G.~E.
  Duggan, K.~Eswaran, P.-H. Cameron~Chen, Y.~Liu, S.~R. Kalidindi, {\em
  et~al.}, ``Chest radiograph interpretation with deep learning models:
  assessment with radiologist-adjudicated reference standards and
  population-adjusted evaluation,'' {\em Radiology}, vol.~294, no.~2,
  pp.~421--431, 2020.

\bibitem{johnson2019mimic}
A.~E. Johnson, T.~J. Pollard, N.~R. Greenbaum, M.~P. Lungren, C.-y. Deng,
  Y.~Peng, Z.~Lu, R.~G. Mark, S.~J. Berkowitz, and S.~Horng, ``Mimic-cxr-jpg, a
  large publicly available database of labeled chest radiographs,'' {\em arXiv
  preprint arXiv:1901.07042}, 2019.

\bibitem{bustos2020padchest}
A.~Bustos, A.~Pertusa, J.-M. Salinas, and M.~de~la Iglesia-Vay{\'a},
  ``Padchest: A large chest x-ray image dataset with multi-label annotated
  reports,'' {\em Medical Image Analysis}, p.~101797, 2020.

\bibitem{demner2016preparing}
D.~Demner-Fushman, M.~D. Kohli, M.~B. Rosenman, S.~E. Shooshan, L.~Rodriguez,
  S.~Antani, G.~R. Thoma, and C.~J. McDonald, ``Preparing a collection of
  radiology examinations for distribution and retrieval,'' {\em Journal of the
  American Medical Informatics Association}, vol.~23, no.~2, pp.~304--310,
  2016.

\bibitem{raghu2019transfusion}
M.~Raghu, C.~Zhang, J.~Kleinberg, and S.~Bengio, ``Transfusion: Understanding
  transfer learning for medical imaging,'' in {\em Advances in neural
  information processing systems}, pp.~3347--3357, 2019.

\bibitem{huang2017densely}
G.~Huang, Z.~Liu, L.~Van Der~Maaten, and K.~Q. Weinberger, ``Densely connected
  convolutional networks,'' in {\em Proceedings of the IEEE conference on
  computer vision and pattern recognition}, pp.~4700--4708, 2017.

\bibitem{islam2020review}
M.~Islam, F.~Karray, R.~Alhajj, J.~Zeng, {\em et~al.}, ``A review on deep
  learning techniques for the diagnosis of novel coronavirus (covid-19),'' {\em
  arXiv preprint arXiv:2008.04815}, 2020.

\bibitem{guo2017calibration}
C.~Guo, G.~Pleiss, Y.~Sun, and K.~Q. Weinberger, ``On calibration of modern
  neural networks,'' {\em arXiv preprint arXiv:1706.04599}, 2017.

\end{thebibliography}
\end{document}